\documentclass[showpacs,preprintnumbers,amsmath,amssymb]{revtex4}
\usepackage{graphicx}
\usepackage{epsfig}
\usepackage{dcolumn}
\usepackage{bm}

\newcommand{\be}{\begin{equation}}
\newcommand{\ee}{\end{equation}}
\newcommand{\ba}{\begin{array}}
\newcommand{\ea}{\end{array}}

\begin{document}

\title{Dual multifractal spectra} 

\author{St\'ephane Roux$^{(1)}$ and Mogens H. Jensen$^{(2)}$}

\affiliation{(1): Laboratoire Surface du Verre et Interfaces, UMR
CNRS/Saint-Gobain,\\
39 quai Lucien lefranc, 93303 Aubervilliers cedex, France.\\
(2): Niels Bohr Institute, Blegdamsvej 17, DK-2100 Copenhagen \O, Denmark.}

\date{\today}

\begin{abstract}
 The multifractal formalism characterizes the
scaling properties of a physical density $\rho$ as a function of
the distance $L$.  To each singularity $\alpha$ of the field is
attributed a fractal dimension for its support $f(\alpha)$. An
alternative  representation has been proposed by Jensen
\cite{jensen} considering the distribution of distances associated
to a fixed mass.  Computing these spectra for a multifractal
Cantor set, it is shown that these two approaches are dual to each
other, and that both spectra as well as the moment scaling
exponents are simply related. We apply the same inversion
formalism to exponents obtained for turbulent statistics in the
GOY shell model and observe that the same duality relation holds
here.
\end{abstract}

\maketitle

\section{Introduction}

Initially motivated by the statistical characterization of
velocity fluctuations in turbulence \cite{turb}, the
multifractal \cite{multifrac1,multifract} formalism has been shown to be a
powerful way of analyzing a large body of different problems.  It
provides a simple and elegant way of performing a ``dimensional
analysis'' of singular fields. In turbulence, this approach has
been applied to the fluctuations of the velocity field, and
deviation from the simple (monofractal) Kolmogorov \cite{kolmo}
scaling of moments of different orders has been observed both
experimentally \cite{multexp1,multexp2,multexp4,multexp3} and
numerically \cite{paladin,gotoh}. It has been used also to characterize the
growth probabilities of Diffusion-Limited Aggregation \cite{DLA}.
Random resistor networks \cite{RRN1,RRN2} at the onset of
percolation have also been studied using this formalism. Extension
to damage and fracture models \cite{damage} have been proposed.

In those examples a local physical quantity $m$ --- referred to as
a ``mass'' in the following for concreteness ---  is distributed
in space (or time) and the formalism allows to characterize the
statistical distribution of this quantity, or equivalently its
moment of any order, as function of the system size $L$ (or time
interval) over which it is considered.  The field is decomposed,
according to its singularities, $\alpha$, into a continuous set of
fractal supports. The corresponding fractal dimension $f(\alpha)$
as a function of the singularity $\alpha$ of the field is the
multifractal spectrum.  Hence, the number $n(m)$ of elements of
mass $m$ such that
 \be
 m\sim L^\alpha
 \ee
 scales as
 \be
 n(m)\sim L^{f(\alpha)}
 \ee

From this function, the scaling of any statistical moment of the
field can be computed. Defining the moment of order $q$, $M_q$,
and its scaling with the system size as
 \be
 M_q=\sum_i m_i^q\sim L^{\tau(q)}
 \ee
we can relate the scaling exponents $\tau(q)$ to the multifractal
spectrum, through a simple Legendre transform \cite{multifract}
 \be\left\{\ba{l}
 q=-f'(\alpha)\\
 \tau(q)=q\alpha+f(\alpha)
 \ea\right.\ee

Recently, Jensen\cite{jensen}  proposed to consider an alternative
approach to characterize the same fields.  Instead of studying the
statistical distribution of mass $m(L)$ over a fixed distance $L$,
he proposed to consider the distribution of distances $L(m)$ such
that a fixed mass $m$ is contained in each subset. From the
initial description $\langle m(L)^q\rangle\propto L^{\tau(q)}$ a
naive expectation would have been that $\langle L(m)^\tau(q)
\rangle\propto m^q$. However, considering the GOY
model\cite{goy1,goy2} (as a toy-model for turbulence), it was
shown that the latter expectation was violated \cite{jensen}.  Instead a
different scaling was observed
  \be
  \langle L(m)^p\propto m^{\theta(p)}
  \ee
  But apparently, the series of exponent $\theta(p)$ seemed
unrelated to the $\tau(q)$.  This unexpected feature suggested to
use this new scaling as a complementary statistical property of
turbulence.

In the following we will consider a simple example of a
multifractal set, using the standard Cantor set construction, but
endowing each interval with a different mass \cite{multifract}.  This simple case
study allows to obtain a direct evaluation of the two multifractal
spectra, as well as the corresponding scaling exponents.  We show
that within this example both approaches are related through
simple duality relations.  We then discuss the applicability of
the previously derived duality relations to the case of
turbulence.  Numerical estimates of the scaling exponents of
length moments for fixed velocity differences are obtained from
She and Levesque \cite{She} proposed form for the velocity moments
scaling exponents. We can apply this formula to obtain the series
of exponents for the inverse statistics. To compare these
``static" data to more realistic dynamical turbulence data we
extract the scaling exponents for forward and inverse statistics
of the GOY shell model. This was already done in \cite{jensen} but
here we extend the analysis to negative values of the moments of
the standard forward structure functions. We then apply the
inversion formula and compare to the exponents obtained by direct
measurements of the inverse structure function. We obtain
quite good results in the comparisons of these data sets as will
be discussed in particular in Section VI. Although it is by no
means a proof, it gives an indication that at least in some cases,
the inversion formula we derive (which has been previously been
derived in other contexts as we discuss) gives a relation between
the exponents of forward and inverse statistics.

\section{Standard multiscaling for the Cantor set}

\begin{figure}
\epsfig{width=9cm,angle=0,file=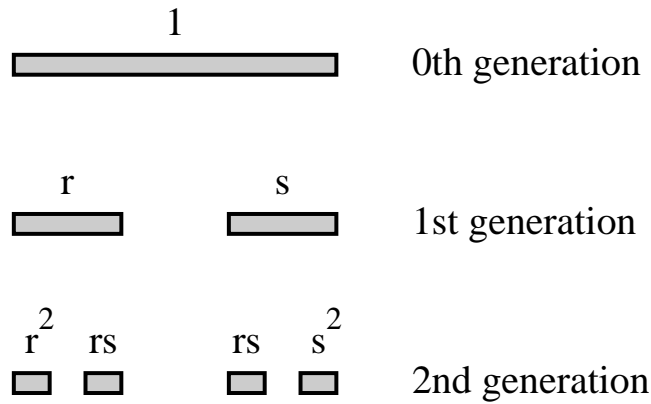}
\caption{\label{fig:cantor} Construction of the multifractal
Cantor set.  Starting from an interval of length $L$ and mass $M$
(generation 0), the first generation is obtained by splitting the
interval into three equal length segments, and dropping the middle
part.  The mass $M$ is distributed into two unequal parts $rM$ and
$sM$ respectively for the left and right interval.  The same
procedure is repeated recursively onto each interval.}
\end{figure}

The interval of length $L$ is split in three equal parts and the
middle one is removed.  The mass $M$ is split in two  unequal
parts, $rM$ and $sM$ such that $r+s=1$.  After $N$ repetitions of
this procedure, we obtain a generation $N$ structure.  The size of
each piece is $\ell=3^{-N} L$.  Its mass is $m=r^is^{N-i}M$, where
$i$ is the number of $r$ choices leading to a specific part. The
number of such intervals carrying the same mass is $n={i \choose
N}$, while the total number of parts is $S=\sum_i {i \choose
N}=2^N$.

We go to the continuum limit and define the real $x$ as $i=xN$.
Using Stirling formula we have
 \be\ba{ll}
 n&=2^{-N} \frac{N^N}{(Nx)^{Nx}(N(1-x))^{N(1-x)}}\\
&= \left({x^{x}(1-x)^{(1-x)}}\right)^{-N}
 \ea\ee

 In order to bridge this computation with the standard
way of defining the multifractal spectrum \cite{multifract}, we introduce
 \be\ba{ll}
 \alpha&=\frac{\log(m/M)}{\log(\ell/L)}\\
 f(\alpha)&=-\frac{\log(n)}{\log(\ell/L)}
 \ea\ee
%
%
 and we define $\alpha_0=-\log(s)/\log(3)$ and
 $\alpha_1=-\log(r)/\log(3)$.  A simple algebra leads to
 \be
 f(\alpha)=\frac{(\alpha_1-\alpha_0)\log(\alpha_1-\alpha_0)
 -(\alpha_1-\alpha  )\log(\alpha_1-\alpha  )
 -(\alpha  -\alpha_0)\log(\alpha  -\alpha_0)}
 {(\alpha_1-\alpha_0)\log(3)}
 \ee
\begin{figure}
\epsfig{width=9cm,angle=0,file=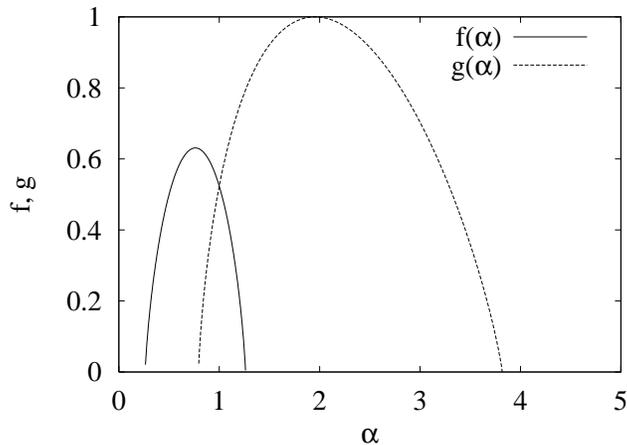}
\caption{\label{fig:falpha} Multifractal spectra for the Cantor
set.  The bold curve is the $f(\alpha)$ function, while the dotted
curve shows the dual spectrum $g(\alpha)$.}
\end{figure}

 \section{Moment scaling}

 The moment of order $r$ of the mass distribution is defined as
 $A_q(N)=\sum_i n(i) (m(i)/M)^q$.  It obeys the recursion
 formula
 \be
 A_q(N)=(r^q+s^q)A_q(N-1)
 \ee
 For the definition of the scaling exponent $\tau(q)$ as
 $A_q(N)\propto (\ell/L)^{-\tau(q)}$ we can write
 \be
 \tau(q)=-\frac{\log(r^q+s^q)}{\log(3)}
 \ee

 One basic property of the multifractal formalism is that the
 scaling exponents $\tau(q)$ can be related to the multifractal
 spectrum through a Legendre transform.  Indeed, the moment can be
 evaluated as $A_q(N)=\sum_\alpha (\ell/L)^{f(\alpha)+q\alpha}$.
 Hence, $\tau(q)=\max_\alpha[f(\alpha)+q\alpha]$.  This defines
 the strength of the singularity $\alpha$ which contributes
 dominantly to the moment of order $q$.
 \be\left\{\ba{ll}
 q&=-f'(\alpha)\\
 \tau(q)&=f(\alpha)+q\alpha
 \ea\right.\ee

 The symmetry property of the Legendre transform allows to express
 the reverse transformation as
 \be\left\{\ba{ll}
 \alpha&=\tau'(q)\\
 f(\alpha)&=\tau(q)-q\alpha
 \ea\right.\ee

 \begin{figure}
\epsfig{width=9cm,angle=0,file=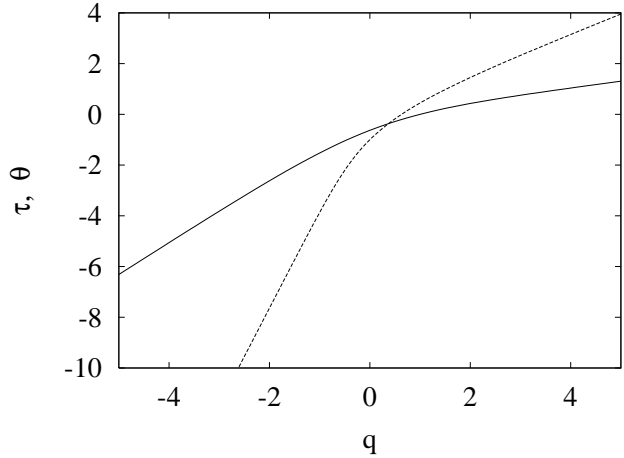}
\caption{\label{fig:scalexpo} Scaling exponents $\tau(q)$ for the
mass as a function of the interval length in bold line, and the
dual scaling exponent $\theta(q)$ in dotted line.}
\end{figure}

 \section{Constant mass ensemble}

 Now we introduce the alternative approach of truncating the
 hierarchical construction at a fixed mass and not fixed
 generation (or length).  The mass $m/M$ is chosen, and thus
 $i$ and $j$ are related by
 \be
 r^i s^j =m/M
 \ee
 The length of such intervals is
 \be
 \frac{\ell}{L}=3^{-(i+j)}
 \ee
 and their number is $n={i\choose i+j}$.

To mimick a similar construction as previously we define
$\ell/L=(m/M)^\beta$, and $n=(m/M)^{g(\beta)}$.  A simple
computation leads to
 \be
 g(\beta)=\frac
 {\beta(\alpha_1-\alpha_0)\log(\beta)
 +\beta(\alpha_1-\alpha_0)\log(\alpha_1-\alpha_0)
 -(1-\beta\alpha_0)\log(1-\beta\alpha_0)
 -(\beta\alpha_1-1)\log(\beta\alpha_1-1)}
 {(\alpha_1-\alpha_0)\log(3)}
 \ee
 Comparison with the original multifractal spectrum shows that
 they are related through
 \be\label{eq:gf}
 g(\beta)=\beta f(1/\beta)
 \ee

In fact this key relation can be simply derived by noting that
 \be\ba{ll}
 n&=(\ell/L)^{f(\alpha)}\\
 &=(m/M)^{f(\log(m/M)/\log(\ell/L))\log(\ell/L)/\log(m/M)}\\
 &=(m/M)^{\beta f(1/\beta)}\\
 &=(m/M)^{g(\beta)}
 \ea\ee
 from which Eq.\ref{eq:gf} results.

\section{Moment scaling}

We introduce similar moments in the dual ensemble $B_p=\sum
n(\ell/L) (\ell/L)^p$ and define their scaling exponents
$\theta(p)$ as $B_p\propto (m/M)^{\theta(p)}$. As for the primal
ensemble, the scaling exponents $\theta(p)$ are related to the
multifractal spectrum $g(\beta)$ through a Legendre transform
 \be\left\{\ba{ll}
 p&=-g'(\beta)\\
 \theta(p)&=g(\beta)+p\beta
 \ea\right.\ee
 and reciprocally
 \be\left\{\ba{ll}
 \beta &=\theta'(p)\\
 g(\beta)&=\theta(p)-p\beta
 \ea\right.\ee

Let us now use the duality relation \ref{eq:gf} to relate the two
series of scaling exponents.
 \be\ba{ll}
 p&=-g'(\beta)\\
 &= -f(1/\beta)+f'(1/\beta)/\beta\\
 &= -f(\alpha)+f'(\alpha)\alpha\\
 &=-\tau(q)+r\alpha-q\alpha\\
 &=-\tau(q)
 \ea\ee
 and
 \be\ba{ll}
 \theta(p)&=g(\beta)+p\beta\\
 &=(f(\alpha)+p)/\alpha\\
 &= (\tau(q)-q\alpha-\tau(q))/\alpha\\
 &=-q
 \ea\ee
Thus the two functions $\tau$ and $\theta$ are related by a mere
inversion up to sign reversals (see also \cite{hastings}),
 \be
 -\theta(-\tau(q))=q
\label{inv}
 \ee
One can also visualize the above result by noting that the graph
of $\theta(p)$ is obtained from that of $\tau(q)$ through a simple
symmetry with respect to the line passing through the origin and
of direction $(-1,1)$.

Therefore, contrary to what was initially proposed, the two series
of exponents are not independent.  They are linked by a duality
relation, as the two multifractal spectra $f$ and $g$.

\section{Application to turbulence statistics}

In order to test the relevance of the above analysis to a more
physical application than the Cantor set, we resort to the
framework of turbulence which was the initial context of the
suggestion of these dual quantities. In this context, the local
physical quantity of interest is the velocity fluctuation $\Delta
u$ (instead of the mass in the above example) studied over a
distance $r$ which plays the role of the local scale $\ell$.  The
early suggestion by Kolmogorov of the scaling moments
$\tau(q)=q/3$ was shown to break down due to intermittency thus
defining a non-trivial series $\tau(q)$ which has resisted all
theoretical attempts to compute them up to now. Nevertheless,
semi-empirical formulas have been proposed which accounts rather
precisely for the numerical values of these exponents as
determined experimentally.  In particular, the She and Levesque
\cite{She} formula appears as an accurate fit.  They proposed
 \be
 \tau(q)=q/9+2\left(1-(2/3)^{q/3}\right)
 \ee
Table \ref{tab:expo} gives the corresponding $\tau$ exponents for
selected positive moments. In order to extract the corresponding
$\theta$ exponents for the dual statistics, we apply the inversion
formula (\ref{inv}). Indeed, we then need to employ the
She-Levesque formula for negative moments. From the derivation of
this formula, it is not obvious that such extension is allowed by
the assumptions made by She and Levesque. Nevertheless, we take
the liberty to continue the formula to negative moments and obtain
the list of exponents $\theta$ listed in Table I.

In order to compare these results with data obtained from numerical simulations
of turbulence models, and even more importantly to test the inversion
formula (\ref{inv}),
we turn to dynamical turbulence generations by shell models \cite{paladin}.
This kind of data were already applied in the paper by one of us which
proposed the inverse structure functions \cite{jensen}. In turbulence theory,
it is well know that scaling behavior of velocity field ${\bf u} ({\bf x}, t)$ and
the understanding of intermittency effects in fully developed turbulence
is described in terms of standard structure functions defined as
\begin{equation}
 \label{vel}
\langle  \Delta  u_{\bf x} (\ell)^q \rangle \sim \ell^{\tau_q}
\end{equation}
where the difference is
\begin{equation}
\label{ua}
\Delta  u_{\bf x} (\ell) ~=~ {\bf u} ({\bf x + r}) - {\bf u} ({\bf x})~~,~~
 \ell = | {\bf r} |
\end{equation}
The average in Eq.(\ref{vel}) is over space and time. We have
assumed full isotropy of the velocity field. The set of exponents
$\tau_q$ forms a multiscaling spectrum \cite{PV}.

\begin{figure}
\epsfig{width=9cm,angle=0,file=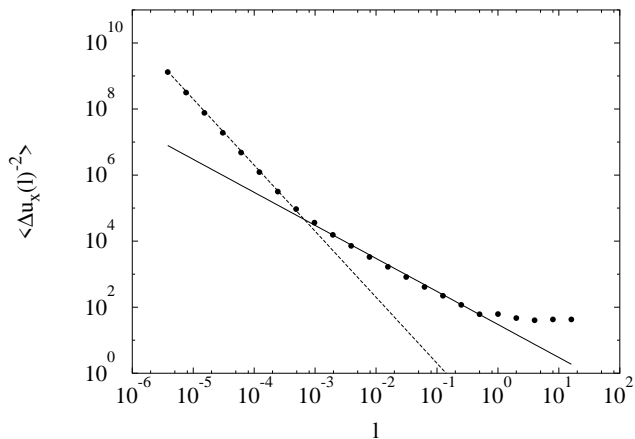} \caption{\label{fig4}A
plot of the ordinary structure function from shell model data with moment $q=-2$, i.e.
$\langle (\Delta  u_{\bf x}(\ell) )^q \rangle$ versus the length
scale $\ell$. Note the three different regimes: The small scale
regime of the smooth behavior; the inertial scaling regime and
outer cut-off. }
\end{figure}

The corresponding dual structure functions is defined by
considering the following quantities
 \begin{equation}\label{ds}
 \langle  \ell (\Delta  u_{\bf x} )^q \rangle \sim | \Delta
 u_{\bf x}|^{\theta_q}
 \end{equation}
where the difference $\Delta u_{\bf x}$ is again defined as in Eq.
(\ref{ua}) and $\ell (\Delta u_{\bf x})$ is understood as the {\it
minimal} distance in $\bf r$, measured from $\bf x$, for which the
velocity difference exceeds the value $\Delta u_{\bf x}$ \cite{jensen}. In other
words, we fix a certain set of values of the velocity difference
$\Delta u_{\bf x}$. Starting out from the point $\bf x$, we
monitor the distances $\ell (\Delta u_{\bf x})$ where the velocity
differences are equal to the prescribed values. Performing an
average over space and time the inverted structure functions Eq.
(\ref{ds}) are obtained.

The turbulence data are obtained from simulations of the GOY shell
model \cite{goy1,goy2,paladin}. This model is a rough approximation to the Navier-Stokes
equations and is formulated on a discrete set of $k$-values,
$k_n=r^n$. We use the standard value $r =2$. In term of a complex
Fourier mode, $u_n$, of the velocity field the model reads
\begin{eqnarray}
\label{un}
(\frac{d}{ dt}+\nu k_n^2 ) \ u_n \ & = &
 i \,k_n (a_n \,   u^*_{n+1} u^*_{n+2} \, + \, \frac{b_n}{2}
u^*_{n-1} u^*_{n+1} \, + \, \nonumber \\
& & \frac{c_n }{4} \,   u^*_{n-1} u^*_{n-2})  \ + \ f \delta_{n,4},
\end{eqnarray}
with boundary conditions $b_1=b_N=c_1=c_2=a_{N-1}=a_N=0$.
$f$  is an external, constant forcing, here on the forth mode.
The  coefficients of the non-linear terms  must follow the relation
$a_n+b_{n+1}+c_{n+2}=0$ in order  to satisfy the conservation  of  energy,
$E = \sum_n |u_n|^2$, when $f=\nu = 0$.
The constraints still leave a free parameter $\epsilon$ so that
 one can set
$ a_n=1,\ b_{n+1}=-\epsilon,\ c_{n+2}=-(1-\epsilon)$ \cite{bif}.
As observed in \cite{kada1}, one obtains the canonical value
$\epsilon= 1/2$, if helicity conservation is also demanded. The
set (\ref{un}) of $N$ coupled ordinary differential equations can
be numerically integrated by standard techniques. We have used
standard parameters in this paper $N = 27, \nu = 10^{-9}, k_0 =
0.05$, and $f = 5 \cdot 10^{-3}$.

\begin{figure}
\epsfig{width=9cm,angle=0,file=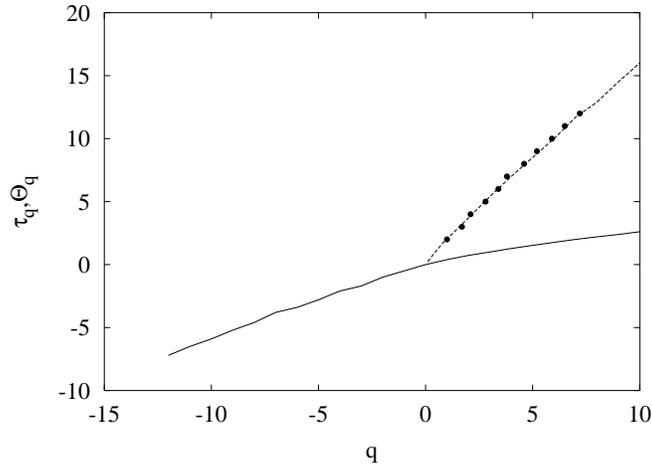}
\caption{\label{fig5}Scaling exponents obtained from simulations
of the GOY shell model. The full curve gives the forward structure
function exponents, $\tau_q$. The dashed line connects the
exponents $\theta_q$ for the inverted structure functions. The
black dot are the results obtained by applying the inversion
formula Eq. \ref{inv} and thus transferring the data for negative
q's of $\tau_q$ to positive q's for $\theta_q$. }
\end{figure}

The structure functions exponents $\tau_q$ are shown in Fig.
\ref{fig5} for integer moments in the interval $q \in [-10;12]$. A
line connects the points in order to guide the eye. The associated
exponents $\theta (q)$ for the inverse structure functions are
also shown in Fig. \ref{fig5} for moments in the interval $q \in
[0;12]$, and are connected by a dotted line. It is possible to
extract the exponents $\tau_q$ to reasonable accuracy for negative
moments $-q$ although the quality of the scaling gradually
decreases with the value of $q$. As an example we show in Fig.
\ref{fig4} the behavior of $ \langle  \ell (\Delta  u_{\bf x}
)^{-2}\rangle$, corresponding to $q=-2$. We observe three distinct
regimes: the small scales referring to the trivial smooth regime,
the ``inertial'' scaling regime and the cut-off regime at large
scales. For increasing value of negative moments, the point where
$ \langle \ell (\Delta  u_{\bf x} )^q \rangle$ are small will be
enhanced.  This is a very important point to be considered for the
analysis of experimental data.  Indeed, the lower cut-off of the
inertial, increasing with large negative values of
$q$ \cite{frischver}, may render the analysis of experimental data quite difficult.
Using an extended self-similarity procedure (studying one moment
against another one rather than as a function of the velocity
difference) seem to provide better results for the inverse exponents, but
since the cut-off effect is physical and not a measurement
artifact, the exploitation of the data may lead to apparent
contradictions \cite{wvdw}.

It is to be noted that from duality, we may convert dual positive
order moment to direct negative order moment.  The latter may be
very difficult to estimate experimentally because of the possible
occurrence of arbitrarily small velocity differences over a given
distance (or time using Taylor's hypothesis).  Therefore, the
duality relation may be exploited to obtain data which would be
inaccessible otherwise.  This procedure however relies on the
applicability of this duality to experimental turbulence, which
could not be tested directly if the direct moments cannot be
computed.  Nevertheless the present test of the analysis, using the
shell model, constitutes an encouraging argument to proceed in this
direction.

Armed with these data sets we are ready to check the validity of
the inversion formula (\ref{inv}) which by a simple inversion
reads
\begin{equation}
-\tau(q) = \theta^{-1} (-q)
\label{b}
\end{equation}

Using the relation (\ref{inv}) directly by inserting $\tau(q)$
for $q \in [-10;-1]$ we obtain by linear extrapolation
the data for $\theta_q$ shown in Table II.
The results are also shown as black dots in Fig. \ref{fig5}.
Using the inverted
relation (\ref{b}), we can compare the value of $-\tau_q$ with
the values of $\theta^{-1} (-q)$, and these are also listed in Table II.
Indeed, there is a very good correspondence between the values
supporting the suggestion that the inversion formula is valid
for the shell model turbulence data. This might be somewhat
surprising as we do indeed perform a very different statistics.

An inversion formula similar to the one derived in this paper has
been proposed for the turbulence spectra \cite{massimo} and
has been applied to multiaffine fields in \cite{luca}. Recently, the
inverse statistics has been applied to two-dimensional turbulence
with the very interesting result that the inverse statistics of a smooth signal
shows non-trivial behavior \cite{2D}.
Hastings has also derived a similar formula for Laplacian random
walks in the very different context of diffusion limited
aggregation \cite{hastings}. By using iterated conformal mappings
Hastings obtain the exact multifractal spectra of the harmonic
measure and derive the inversion formula for the $f(\alpha)$
spectrum.

One might express a general worry that in the integrals for the
statistical averages one integrates over the same singularity
structure, both to obtain the standard and inverse (i.e. dual) structure functions.
That is to say that dominating terms of the integrals
(and thus the saddle point) come from the same singular structures
even though one might argue that for turbulence, the velocity
singularities are important for the forward structure functions
whereas the laminar regions are important for the dual structure
functions.  However, as mentioned earlier, besides the intrinsic
limitations due to the evolution of the cross-over length scale
between laminar and inertial regimes with the moment order, when
both tools are used to analyze the inertial regime, we are
characterizing the same multifractal object for which the
correspondence is expected to hold.  Indeed the shell model used
to validate the procedure is not deprived of such singularities
(looking like shock waves in that case).

\section{Conclusions}

The alternative approach to the standard multifractal spectrum and
scaling exponents of different moment orders, which was proposed by
interchanging the role of the physical quantity of interest and
the length (or time) scale, has been examined in the case of a simple
multifractal Cantor set.  This example shows that the two spectra
are simply related, and that the scaling exponents of the length
moments can be related to the usual series of scaling exponents.

Based on this correspondence, we tested the application of this
alternative approach to turbulence using data obtained
from shell model calculations. A good agreement was
found, thus suggesting that the above duality relation could be
extended to more general cases.

\begin{acknowledgments}
We wish to acknowledge fruitful discussions and exchanges
with Massimo Vergassola and Willem van der Water.
\end{acknowledgments}

\begin{table}
\caption{\label{tab:expo} Value of some $\tau(q)$ and $\theta(q)$
exponents for moments of order $q$ based on numerical simulations
of the GOY shell model \cite{goy1,goy2,paladin}. These data are as presented
in Ref. \cite{jensen}. For comparison, we present the corresponding
series of exponents based on She and Levesque
formula  (index $SL$). To obtain the series of the inverse
exponents $\theta_{SL}(q)$ we have invoked the inversion formula (\ref{inv}).}
\begin{ruledtabular}
\begin{tabular}{ccccc}
  $q$     & $\tau_{GOY}(q)$& $\theta_{GOY}(q)$ & $\tau_{SL}(q)$ & $\theta_{SL}(q)$  \\
\hline
  0.      &  0.00  &  0.00  &  0.000 &  0.00  \\
  0.2     &  0.08  &  0.45  & 0.076  &  0.51  \\
  0.4     &  0.15  &  0.89  &  0.150 &  1.00  \\
  0.6     &  0.22  &  1.3   &  0.222 &  1.46  \\
  0.8     &  0.29  &  1.7   &  0.294 &  1.91  \\
  1.      &  0.39  &  2.04  &  0.364 &  2.33  \\
  2.      &  0.73  &  3.7   &  0.696 &  4.21  \\
  3.      &  1.00  &  5.4   &  1.00  &  5.77  \\
  4.      &  1.28  &  7.0   &  1.28  &  7.09  \\
  5.      &  1.53  &  8.5   &  1.54  &  8.23  \\
  6.      &  1.77  &  10.0  &  1.78  &  9.24  \\
  7.      &  2.00  &  11.7  &  2.00  & 10.14  \\
  8.      &  2.20  &  12.9  &  2.21  & 10.95  \\
  9.      &  2.39  &  14.5  &  2.41  & 11.68  \\
 10.      &  2.61  &  16.0  &  2.59  & 12.36  \\
\end{tabular}
\end{ruledtabular}
\end{table}

\begin{table}
\caption{\label{tab:expo1} Value of exponents obtained from
simulations of the GOY model for negative moments in
the interval $q \in [-12;-2]$. Shown are the values
of $- \tau(q)$ which according to the inversion formula
should be compared to $ \theta^{-1} (-q)$. Last row are
values of $\theta(-\tau(q))$ which should be compared to $-q$.}
\begin{ruledtabular}
\begin{tabular}{cccc}
  $q$     & $- \tau(q)$ & $ \theta^{-1} (-q)$ & $\theta(-\tau(q))$ \\
\hline
  -2      &  1.0  &  0.98  & 2.04   \\
  -3      &  1.7  &  1.59  & 3.1    \\
  -4      &  2.1  &  2.17  & 3.9    \\
  -5      &  2.8  &  2.76  & 5.06   \\
  -6      &  3.4  &  3.38  & 6.04   \\
  -7      &  3.8  &  4.0   & 6.7    \\
  -8      &  4.6  &  4.66  & 7.9    \\
  -9      &  5.2  &  5.34  & 8.8    \\
  -10     &  5.9  &  6.0   & 9.85   \\
  -11     &  6.5  &  6.59  & 10.85  \\
  -12     &  7.2  &  7.25  & 11.9   \\
\end{tabular}
\end{ruledtabular}
\end{table}

\end{document}